\documentstyle[preprint,aps,epsfig]{revtex}

\begin{document}

\title{Scattering Theory of Kondo Mirages and Observation of Single
  Kondo Atom Phase Shift}
\author{Gregory A. Fiete$^{1}$, Jesse S.
Hersch$^{1}$ and Eric J. Heller$^{1,2}$}

\address{ $^1$Department of Physics, Harvard University, Cambridge,
Massachusetts 02138\\ $^2$Department of Chemistry and Chemical Biology, Harvard
University, Cambridge, Massachusetts 02138\\ }

\author{H.C. Manoharan, C.P. Lutz, and D. M. Eigler}

\address{ IBM Research Division, Almaden Research Center, 650 Harry
  Road, San Jose, California 95120} 


\maketitle

\begin{abstract} We explain the origin of the Kondo mirage seen in
  recent quantum corral Scanning Tunneling Microscope (STM)
  experiments with a scattering theory of electrons on the 
  surfaces of metals.  Our theory combined with experimental data 
  provides the first
  direct observation of a single Kondo atom phase shift.  The Kondo
  mirage observed at the empty focus of an elliptical quantum corral is shown
  to arise from multiple electron bounces off the corral wall adatoms 
in a
  manner analagous to the formation of a real image in optics.  We
  demonstrate our theory with direct quantitive comparision to 
  experimental data.
\\ \\ PACS numbers: 73.20.At, 72.10.Fk, 72.15.Qm, 3.65.Nk \\
\end{abstract}

The Kondo effect is a fascinating, many-body quantum phenomenon
occuring at low temperatures, whereby
a magnetic impurity in metallic bulk or on a metallic surface has its local
moment screened by a cloud of conduction electrons, called the Kondo
cloud, that forms in its vicinity\cite{hewson}.  Until recently, only bulk
measurements were possible on Kondo systems, leaving isolated Kondo
atoms and their scattering phase shifts unstudied.  The atomic
resolution of the STM now makes the study of single, isolated Kondo atoms
possible \cite{hari,li,madhavan}.  Kondo adatoms are identified by a
sharp ($\sim 10$ meV wide)
feature in the differential tunneling conductance ($dI/dV$) as a
function of STM tip bias with respect to the surface of the material.
The feature typically appears as a dip near the Fermi energy
($E_F$) in $dI/dV$, is localized to within $\sim$ 10 \AA\ of a Kondo
impurity and is only observed at 
low temperatures ($\sim 4$ K). If the temperature of the substrate 
becomes to high ($\sim 100$ K), the correlations between the impurity
spin and the conductions electrons are broken and the Kondo signature 
disappears.

Landmark STM experiments have recently discovered the remarkable
fact that when a Kondo atom is placed at one focus of a
properly sized empty elliptical quantum corral built from Kondo
adatoms, a ``mirage'' of the Kondo
feature is cast to the opposite focus \cite{hari} more than 70 \AA\
away.  Since the Kondo effect arises from electron correlations,
what does the Kondo feature at the unoccupied focus imply about local
electron correlations there?  An important feature of the mirage
experiments is that they were done on the surface of Cu(111) which 
is known to have surfaces states which act as a 2-dimensional electron
gas.  The central result of this paper is to
present a theory of the Kondo mirage based on electrons in this
surface state scattering to
infinite order from all Kondo atoms--both the Kondo atoms that make 
up the walls of the corral and the Kondo atom that sits at the
focus. The theory is valid for all arrangements of adatoms, whether
arranged into a corral shape or any other arbitrary structure
including ``open'' structures.    
Electron scattering on the surface can be directly related to 
the $dI/dV$ of the STM
measurements.  Thus, by solving the electron scattering problem on the
surface in the presense of impurities we can compute the STM conductance\cite{rick}.

The basic Hamiltonian in the Kondo problem is given by 
\begin{equation}
H=\sum_{{\bf k}\sigma}\epsilon_k c^\dagger_{{\bf k}\sigma}c_{{\bf
    k}\sigma} +{J\over N}\sum_{{\bf k,k'}\sigma \sigma'}{\bf S} \bullet
    {\bf s}_{\sigma \sigma'} c^\dagger_{{\bf k}\sigma}c_{{\bf k'}\sigma'}
\end{equation}
where $\epsilon_k$ is the dispersion relation (given by band structure
calculations or experiment) of the conduction
electrons, $c^\dagger_{{\bf k}\sigma}$ is an operator that creates an
electron in the state ${\bf k}$ with spin $\sigma$ and $c_{{\bf k}\sigma}$
destroys an electron in the state ${\bf k}$ with spin $\sigma$.  This first
term describes the energy of free conduction electrons.  The second
term describes spin-flip scattering processes where a conduction
electron flips its spin and the impurity changes its spin in response 
during a scattering event.  $N$ is the number of sites in the lattice,
$J$ is the coupling constant, ${\bf S}$ represents the impurity spin and ${\bf
  s_{\sigma \sigma'}}$ represents the conduction electron spin.  It is
well known that the
  Kondo effect occurs when $J>0$ \cite{hewson} and the
  low energy properties of this 
Hamiltonian are described by an
  effective Hamiltonian with $J\to\infty$ so that the conduction
  electrons become ``locked'' into an anti-aligned state with the impurity
  moment.  Spin-flip process are thus frozen out.  We do not attempt
  to solve this Hamiltonian, but assume that we are working in the low
  temperature regime where spin flips are frozen out as 
experiment suggests from the
  feature in $dI/dV$ described earlier.

In the Fermi liquid description 
of Kondo impurities below their Kondo temperature, $T_K$, the
impurities may be characterized by a scattering phase shift that they
impart to impinging electrons (quasi-particles) \cite{nozieres}.  For
$T<<T_K$ Kondo impurities act as potential scatterers with all the
many-body physics appearing as an energy scale in the (resonant) 
scattering phase shift\cite{hewson,nozieres}.  We
apply our theory to experimental data to extract the phase shift
with full energy dependence of a single Kondo atom and report its
value for the first time ever and present calculations that show
beautiful {\it quantitative} agreement with experiment.  Our theory
clarifies what the STM does and does not measure in mirage experiments
regarding local correlations of electrons.  In particular, our theory
suggests that the mirage experiments do not probe local
electron-electron correlations because the Kondo correlations 
only appear in the theory as an energy scale, $T_K$, in the scattering
phase shift (or alternatively, in the density of states). 

We emphasize that our theory is phenomenological in that we do not
attempt to compute the tunneling conductance from first principles on
top of an atom as was done in\cite{kawasaka,schiller}.  Our theory is based on
electron scattering which is by its nature asymptotic--an
electron comes in, strikes a Kondo adatom, and leaves with a phase
shift.  Therefore, our theory only makes accurate quantitative
predictions when the STM tip is more than 7 \AA\ away laterally from a
Kondo adatom on the surface.  This is precisely the region where we can
make accurate predictions about the quantum mirage with our theory.
When the STM tip is within 7 \AA\ laterally of an impurity there is accumulated
screening charge and orbital density present that is not accounted for
in our theory and thus the theory has no predictive power there.  

When an STM tip is biased negatively with respect to the surface of a
metal,
such as Cu(111), electrons can tunnel from the tip onto the surface,
creating
a region of enhanced electron amplitude under the tip which travels
outwards as a wave on the surface.  This electron wave may encounter
surface
defects such as adatoms or step edges, which cause scattering.  Part of
the
scattered wave returns to the STM tip where, depending on the relative
phase
of the outgoing and incoming amplitudes, it interferes constructively or
destructively with the outgoing wave. This interference leads to
fluctuations in the STM current as the tip is moved over the surface.
The STM, then, is sensitive to the {\it coherent} amplitude that
is reflected from defects on the surface. 
Heller {\it et al.}\cite{rick} showed that because the
Fermi wavelength
of the surface state electrons on Cu(111) is much larger than the size of
adatoms and
because the adatoms are separated by a distance large  compared to their
size, it is permissible to use a multiple s-wave scattering expansion to
calculate the electron amplitude on the surface.   In this
picture, the scattered electron wave, and therefore the STM signal, are
determined by a single quantity: the s-wave phase shift of the
scattered wave,
$\delta_o(\epsilon)$, which will typically by complex. The imaginary part of the phase
shift represents ``absorption'' (incoherent scattering of electrons) 
by the adatoms which tend to couple surface states to
bulk states\cite{rick,hormandinger,crampin} resulting in a
loss of electrons from the surface states.  The loss of electrons
from the surface states at impurities has measurable consequences 
in STM experiments and is related to the attenuation of the mirage at
the empty focus of an ellipse relative to the Kondo feature at the
opposite, occupied focus. 

To compute $dI/dV$ from a scattering calculation, we follow the method
of Heller {\it et al}\cite{rick}.  For s-wave scattering from a single adatom in two dimensions, the wave
function is \cite{landau}, $\psi({\bf r}) \to
\phi({\bf r})+  f{e^{i(kr-{\pi\over4})}\over \sqrt{r}}$,
where the scattering amplitude, $f$, is
$f={e^{2i\delta_o(\epsilon)}-1\over \sqrt{2\pi k}}$. $\phi({\bf r})$ is the amplitude of the  circular electron wave emanating
from the STM tip. There is no angular dependence of $f$ in the s-wave
approximation.  $k={2\pi\over \lambda}$, where $\lambda$, the
wavelength of the electrons on Cu(111), is
29.5\AA. $\epsilon(k)={\hbar^2 k^2\over 2m^*}$ is known
from the surface state electron dispersion relation of Cu(111) which 
has an effective mass of $m^*=0.38m_e$.   Once
$\delta_o(\epsilon)$ is determined, the surface scattering  can be
computed.
In the limit of small bias voltages and low temperature\cite{tersoff},
\begin{equation} {dI\over dV} \propto
LDOS(\vec{r},\epsilon)=\sum_\nu|\psi_\nu(\vec{r})|^2\delta(E_\nu-\epsilon),\end{equation} where $\epsilon$ is the energy determined by the bias
voltage V and the Fermi energy E$_F$, via $\epsilon$=E$_F$+$e$V; $\nu$
labels the scattering eigenstates in the presence of the adatom.
When several or many adatoms are present as in this paper, a multiple
scattering approach \cite{rick} is used to compute
$LDOS(\vec{r},\epsilon)$. By definition the $LDOS$ is related to the
Green's function by
$LDOS(\vec{r},\epsilon)=-{1\over \pi} Im[G_{ret}(\vec{r},\epsilon)]$, where 
\begin{equation}
G_{ret}=G_o+G_oTG_o.
\end{equation}
is the retarded Green's function and $T$ is the T-matrix whose 
dimensions are N by N when there are N
scatterers present.  The T-matrix contains all the information about
the physical positions of the scatters relative to each other and
their corresponding phase
shifts which could be different for each
scatterer. $G_o(\vec{r},\epsilon)$, the free electron's Green's
function in two dimensions. Thus,
$dI/dV$ of the STM may be obtained by solving the scattering problem.
 
Our theory, then, involves the following approximations, assumptions
and limitations: (i) We can characterized the ``target'' (adatom) by a single
parameter, the s-wave phase shift and this must be determined from
experiment or otherwise (ii) The internal degrees of freedom (spin)
of the Kondo adatoms are frozen out at the temperature of the
experiment ($\sim 4$K) so we may use the results of Nozieres\cite{nozieres} to
treat the Kondo atom as a potential scatterer with a phase shift (iii) 
The adatoms are far enough apart so that we may treat
the electron propagation between them as free and that RKKY
interactions are sufficiently weak so that the single-impurity Kondo
physics is not altered (iv) The theory does not
include any non-equilibrium effects and does not treat the charge
density right at an atom correctly.

To make a direct comparison with experiment, we must obtain the phase
shift of the Kondo adatoms.  
We do not have an {\it ab initio} calculation of the phase shift
of a single Co adatom. Rather, we fit the resonant form of the
phase shift, including inelasticity,  and  calculated the multiple
scattering problem with this single atom data.  We emphasize that we
have not blithely ignored the Kondo effect of the wall and focal
adatoms in our phenomenological fit to the scattering phase shift.  We
have used the fact that the spin-flips have been frozen
out to lowest order at the temperatures of the experiment and that the
Kondo effect appears through the energy scale of the resonant spectral 
peak in the density of states near $E_F$\cite{hewson}. 

Since the on-atom electron orbital density is not
accounted for in scattering theory, we used an on-atom fit involving
only a renormalization of the free-space Green's function and a change
in the background phase shift to compute the STM
signal on top of a Kondo adatom\cite{kawasaka,schiller,ujsaghy}.  
This on-atom fit is not part of our theory, but only a means of
setting a reference point between on-atom density not accounted for in
our theory and the electron density anywhere more than 7 \AA\ away
from an atom on the surface which {\it is} accounted for properly in
our theory.  This fit in no way compromises our fundamental result
that the mirage is due to resonantly scattering electrons from the
Kondo atoms of the walls and focus.  It is used only as a method to
determine as accurately as possible the phase shift of the Co on Cu(111).
Determining the phase
shift this way from experimental data constitutes a measurement of 
the single Kondo atom phase shift for the first time.

We find a good fit to the
s-wave scattering phase shift to be 
$\delta_o(\epsilon)=\delta_{bg}+i\delta''+$tan$^{-1}({\epsilon-\epsilon_o
\over \Gamma/2}), \label{phase}$  
where
$\delta_{bg}={\pi\over
4}\pm{\pi\over10}$,  $\delta''={3\over 2}\pm{1\over4}$, $\Gamma=(9 \pm 1)$
meV and $\epsilon_o$ = $E_F$ -1 meV are determined by experiment. 
$\delta_{bg}$ is a background phase shift (possibly due to static
charge screening at the impurity) that controls the resonant line
shape of the adatom scattering cross-section.  
$\delta''$ is the imaginary
part of the phase shift and is a measure of the inelasticity in adatom
scattering. Tan$^{-1}({\epsilon-\epsilon_o\over \Gamma/2})$
reflects resonant scattering due to the presence of Kondo physics.  
A similar form of
phase shift has been derived by Ujsaghy from a more
microscopic point of view in \cite{ujsaghy}.
The narrow spectral peak near $E_F$ leads to
resonant scattering Fermi surface electrons and sets the scale of the
resonance in the phase shift.  It is likely that both bulk and surface
states are participating in the Kondo effect at an adatom, but the STM
signal is dominated by the surface state Kondo effect in the regime of
validity of our theory ($>$ 7 \AA\ away from adatom).

Applying the theory to elliptical corrals results in the images shown in
Fig.~1 and Fig.~2.  The agreement with experiment is excellent.
Our calculation of the tunneling
spectrum at the two foci is compared with experiment in
Fig.~3.  Note that the signal at the unoccupied focus
is attenuated by approximately a factor of 8, both experimentally and
theoretically. 
$\delta''$ is largely responsible for the attenuation of the  mirage at
the unoccupied
focus of the ellipse: the rest of the attenuation comes from flux leaking
through the corral.  The calculated spectroscopy in Fig.~3
most clearly demonstrates that the
Kondo mirage is due to resonant scattering of electrons from the
  adatom at the opposite focus. (Even though the electrons are also
  resonantly scattering from the wall adatoms. 
Our calculations show that the wall atoms' Kondo resonances play no
essential role in the projection of the mirage to the empty focus.
Walls with $\delta=i\infty$\cite{rick} also result in the mirage at the empty
focus provided the focal adatom is treated as a resonant scatterer.
Experimentally the same result is found\cite{hari}.)   

The Kondo mirage results from resonant scattering at the Kondo adatoms
and from the geometry of the quantum corral. Although any ellipse will 
focus rays at the foci, but only certain sized ellipses will give a 
good mirage effect--those
which have large surface state amplitude at the foci when the
scattering problem is calculated and this depends on the relative size
of the ellipse and $\lambda_F$.  Only then will there be appreciable
surface state electron amplitude at the focal adatom to give a
Kondo effect in the surface states of Cu(111).  Our theory predicts
that the Kondo mirage is not restricted to an ellipse or even a
``closed'' structure.  Any time one can construct an arrangement of
adatoms or other defects that lead to a build up of surface state
electron amplitude at two locations within the electron's coherence
length, a mirage can be projected. 

Returning to our original question of whether the mirage reveals any
  information about local correlation, we conclude that it does not. 
There is no explicit information
  about electron correlations in our theory, which gives remarkable
  agreement with experiment in reproducing the Kondo mirage and
  standing wave patterns in the elliptical corral.  We are
  thus led to the conclusion that the mirage at the empty focus of the
  elliptical corral is not a result of electron correlations under the
  tip, but rather resonant scattering of electrons from the focal
  adatom and scattering (resonantly or not) from the adatoms of the
  walls of the corral.  Intuitively this makes sense because the
  electrons tunneling out of the STM are unpolarized so it is not
  possible for the STM to give any direct information about electron
  correlations.  The unpolarized STM only returns an average signal 
of spin up and spin down electrons tunneling into the surface.  

In summary believe that the term ``mirage'' is very apt indeed:
the Kondo effect {\it appears} to be projected to a remote spot, but
the STM is only probing the Kondo focal adatom from afar. The
remote image of the Kondo effect is explained quantitatively by the
refocusing of the s-wave electrons leaving the empty focus and traveling
to the occupied focus after bouncing off the walls.

Is the Kondo effect really projected to the empty focus?  Yes, and no.
One need look no further than the concept of the real image in optics
to understand the sense in which projection occurs.

We acknowledge enlightening
discussions with B.I. Halperin, S. Kehrein, and Y. Oreg.  This 
research was supported by the National Science Foundation under 
Grant No. CHE9610501 and by ITAMP.

While preparing this manuscript we became aware of several papers on a
similar topic: Weissman {\it et al}, cond-mat/0007485; Porras {\it et
  al}, cond-mat/0007445; and Agam {\it et al} cond-mat/0006443.

\vskip .5 in
Fig.~(1).  Topograph standing wave patterns of a Kondo corral.  
Using the scattering theory and phase shifts
described in the text, these STM topograph images were computed using
exact Co adatom positions on Cu(111) at 4 K. 
The agreement between theory ({\bf a}, {\bf c} and {\bf e}) and 
experiment ({\bf b}, {\bf d} and {\bf f}) is remarkable.  All the 
experimental images have been symmetrized by adding the 
image to itself after being reflected about its major axis. 
Topographic images were calculated by numerically integrating
the $LDOS({\vec r},\epsilon)$ over $\epsilon$ for $E_F \leq \epsilon
\leq E_F + 10$ mV.  This corresponds to the topographic images taken
experimentally in {\bf b} and {\bf d} at a bias voltage of 10 mV. {\bf
  e} is the difference of {\bf a} and {\bf c}. {\bf f} is the
difference of {\bf b} and {\bf d}. 

\vskip .5 in
Fig.~(2).  $dI/dV$ Standing wave patterns of a Kondo corral.
Same theory vs. experiment arrangment as in Fig.~(1).  $dI/dV$ measurements
were taken simultaneously with topographic images at a 10 $meV$ bias.
Note that {\bf e} and {\bf f} resemble an eigenstate of the ellipse.  
The ellipse was constructed to have large surface state amplitude at
the two foci.

\vskip .5 in
Fig.~(3).  Tunneling into the focal atom and empty focus: The Mirage.  
Tunneling spectroscopy is calculated (dashed lines)
with the scattering theory and phase shift given in the text at 
the empty focus {\bf a}. Tunneling
spectroscopy at the occupied focus is shown in {\bf b}.  
A constant background slope
has been removed from both the experimental data and the calculation. 
The attenuation of the mirage is
determined by inelasticity in the scattering of electrons at the walls of
the ellipse.  The theoretical signal 5\AA\ away from the empty
focus in {\bf a} is lost in the noise of the experiment.
\pagebreak

\begin{figure}
\epsfig{figure=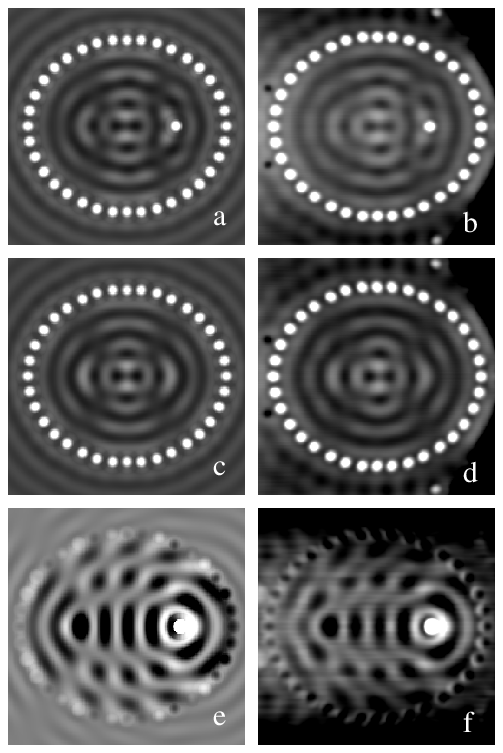,height= 8.0 in}
\label{images_1}
\end{figure}
\pagebreak
\begin{figure}
\epsfig{figure=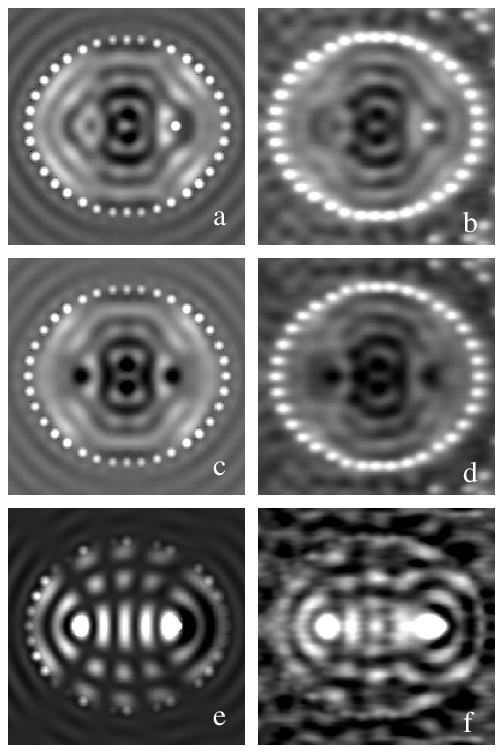,height= 8.0 in}
\label{images_2}
\end{figure}
\pagebreak
\begin{figure} 
\epsfig{figure=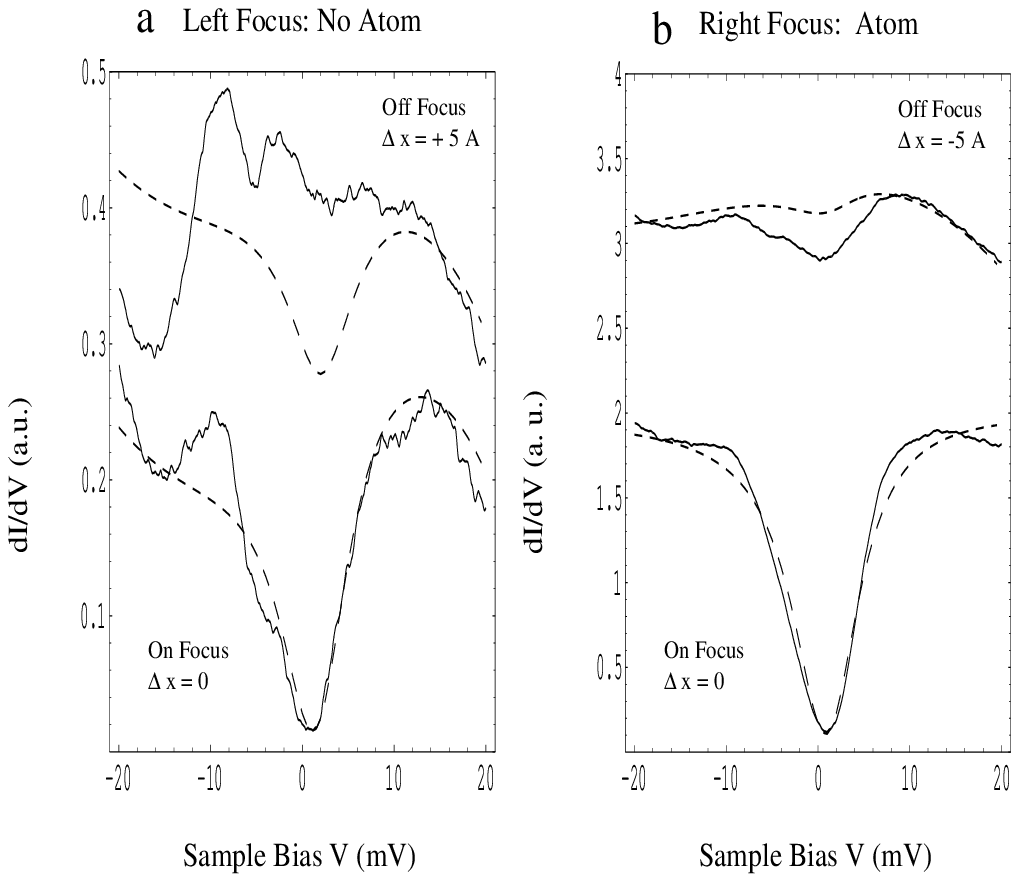, height =4 in}
\label{onatom}
\end{figure}

\end{document}